\documentclass[12pt]{article}
\usepackage[utf8]{inputenc}
\usepackage{amsmath,graphicx}
\usepackage[square,numbers,sort&compress]{natbib}
\usepackage{fullpage}
\usepackage{lineno}
\usepackage{xcolor}
\usepackage{pdflscape}
\usepackage{hyperref}
\usepackage{authblk}

\newcommand{\blue}{}                    % use this for an unmarked copy

\title{Structured methods for parameter inference and uncertainty quantification for mechanistic models in the life sciences}

\author[1]{Michael J. Plank}
\author[2]{Matthew J. Simpson}
\affil[1]{School of Mathematics and Statistics, University of Canterbury, Christchurch, New Zealand}
\affil[2]{School of Mathematical Sciences, Queensland University of Technology, Brisbane, Australia}

\date{}

\begin{document}

\maketitle

\begin{abstract}
Parameter inference and uncertainty quantification are important steps when relating mathematical models to real-world observations, and when estimating uncertainty in model predictions. However, methods for doing this can be computationally expensive, particularly when the number of unknown model parameters is large. The aim of this study is to develop and test an efficient profile likelihood-based method, which takes advantage of the structure of the mathematical model being used. 
We do this by identifying specific parameters that affect model output in a known way, such as a linear scaling. We illustrate the method by applying it to three toy models from different areas of the life sciences: (i) a predator-prey model from ecology; (ii) a compartment-based epidemic model from health sciences; and, (iii) an advection-diffusion-reaction model describing transport of dissolved solutes from environmental science. We show that the new method produces results of comparable accuracy to existing profile likelihood methods, but with substantially fewer evaluations of the forward model.  We conclude that our method could provide a much more efficient approach to parameter inference for models where a structured approach is feasible. Code to apply the new method to user-supplied models and data is provided via a publicly accessible repository.
\end{abstract}


\clearpage

\section*{Introduction}
Parameter inference and uncertainty quantification are important whenever we wish to interpret real-world data, or to make predictions of that data using mathematical models.  This is especially true for modelling applications in the life sciences where data is often scarce and uncertain.   Commonly used methods include tools from both frequentist (e.g. maximum likelihood estimation and profile likelihood) \cite{simpson2021profile,shuttleworth2024empirical} and Bayesian statistics (e.g. Markov chain Monte Carlo methods and approximate Bayesian computation) \cite{sisson2007sequential,toni2009approximate,sunnaaker2013approximate}. These methods can be computationally expensive, particularly for mathematical models with many unknown parameters, resulting in a high-dimensional search space. This has led to a significant body of literature concerned with improving the efficiency of parameter inference methods.

In 2014, Hines et al.~\cite{hines2014} reviewed the application of Bayesian Markov chain Monte Carlo (MCMC) sampling methods for parameter estimation and parameter identifiability for a range of ordinary differential equation (ODE)-based models of chemical and biochemical networks. Their results indicated that parameter non-identifiability can be detected through MCMC chains failing to converge. Similar observations were made by Siekmann et al. \cite{siekmann2012} for a range of continuous-time Markov chain models used in the study of cardiac electrophysiology. In 2020, Simpson et al.~\cite{simpson2020practical} compared MCMC sampling with a profile likelihood-based method for parameter inference and parameter identifiability for a range of nonlinear partial differential equation (PDE)-based models used to interrogate cell biology experiments. The authors studied identifiable and non-identifiable problems and found that, while both the MCMC and profile likelihood-based approaches gave similar results for identifiable models, only the profile likelihood approach provided mechanistic insight for non-identifiable problems. They also found that profile likelihood was approximately an order of magnitude faster to run than MCMC sampling, regardless of whether the problem was identifiable or not.  The speed-up in computation is a consequence of the fact that it is often faster to use numerical optimisation methods compared to sampling methods. Raue et al.~\cite{Raue2013} suggested the sequential use of profile likelihood to constrain prior distributions before applying MCMC sampling in the face of non-identifiability.

While profile likelihood has long been used to assess parameter identifiability and parameter estimation~\cite{Bates1988,Raue2009,Kreutz2013,Kreutz2013b,Ciocanel2023}, Simpson and Maclaren~\cite{simpson2023profile} recently presented a profile likelihood-based workflow covering identifiability, estimation and prediction. The workflow uses computationally efficient optimisation-based methods to estimate the maximum likelihood, and then explores the curvature of the likelihood through a series of univariate profile likelihood functions that target one parameter at a time. This workflow then propagates uncertainty in parameter estimates into model predictions to provide insight into how variability in data leads to uncertainty in model predictions. While the initial presentation of the profile likelihood-based workflow focused on deterministic models, the same ideas can be applied to stochastic models whenever a surrogate likelihood function is available~\cite{simpson2021profile}.

For many mathematical modelling applications, there is little alternative than to either assume a fixed value for a particular parameter, or include it as a target for inference. However, in some cases, there may be parameters whose values are unknown, but whose effect on the model solution is known to be a simple linear scaling or some other known transformation, as we will demonstrate later. In such cases, finding the optimal (likelihood-maximising) values of these parameters is trivial if the other model parameters are known. However, typically there will be other model parameters that are unknown. Performing standard inference procedures on the full set of unknown parameters is inefficient as it fails to make use the simple scaling relationship associated with some of the parameters. 

As a motivating example, \cite{lustig2023modelling} described an epidemiological model that was fitted to data in real-time and used to provide policy advice during the Covid-19 pandemic. The model consisted of a system of several thousand ODEs, and model fitting and uncertainty quantification was done using an approximate Bayesian computation method targeting 11 unknown parameters. This resulted in a computationally intensive problem with a high-dimensional parameter space. Two of the fitted parameters were multiplicative factors on the proportion of model infections that led to hospitalisation and death respectively in each age group and susceptibility class. Adjusting either or both of these parameters while holding the other parameters fixed would linearly scale the time series for expected daily hospital admissions and deaths output by the model. Thus, for a given combination of the other nine fitted parameters, it should be possible to find the optimal values for these two parameters without a costly re-evaluation of the full model.

Here, we propose an new approach, which we term \textit{structured inference}, that exploits the known scaling relationship between certain parameters and the model solution. Our approach is similar to that of Loos et al. \cite{loos2018hierarchical}, who developed a hierarchical parameter inference method for ODE models with additive Gaussian or Laplace noise where some of the parameters, which they termed `scaling parameters', were simply multiplicative factors on expected value of the observed variables. This method was generalised by \cite{schmiester2020efficient} to include offset parameters. Like \cite{loos2018hierarchical,schmiester2020efficient}, our approach recasts an $N$-dimensional optimisation problem as a nested pair of lower-dimensional problems, effectively reducing the dimensionality of the search space. However, our methods go beyond those of \cite{loos2018hierarchical,schmiester2020efficient} in two important aspects: (1) they focus on parameter estimation, whereas we consider parameter estimation, practical identifiability and uncertainty quantification via the profile likelihood; (2) we design the method to be applicable in more general settings, including different classes of mathematical models than just ODE models, more general parameter relationships than just multiplicative scaling, and more general noise models.

We illustrate our method using canonical toy models representing three cases studies drawn from different areas of the life and physical sciences: ecological species interactions; epidemiological dynamics; and pollutant transport and deposition. These examples demonstrate some of the features described under (2) above, e.g. Poisson and negative binomial noise models in the first two examples, and a PDE model with a non-trivial parameter relationship in the third example. In each case study, we use the model structure to identify parameters that have a known scaling effect on model output. We then show that implementing a structured inference approach results in solutions with very similar accuracy, but substantially fewer calls to the forward model solver. This shows that our method could potentially offer significant reduction in computation time when applied to more complex models, which are computationally expensive to solve. Alongside this article, we provide fully documented code for implementing our method, with instructions for how it can be applied to user-supplied models and data.

\section*{Methods}

\subsection*{Unstructured and structured inference}

Suppose a model has $N$ unknown parameters denoted $\theta$. We assume that a likelihood function $L(x,\theta)$ is available for the model for given observed data $x$.  A standard approach to parameter inference is to perform maximum likelihood in the $N$-dimensional parameter space. The maximum likelihood estimate $\theta_\mathrm{MLE}$ for the parameters is the solution of the optimisation problem
\begin{equation} \label{eq:MLE_problem}
\theta_\mathrm{MLE} = \mathrm{argmax}_{\theta} L(x,\theta).
\end{equation}
When solving this problem numerically, each call to the objective function with a given combination of parameters $\theta$ typically requires calculation of the model solution, denoted $y(\theta)$, in order to calculate the likelihood.  This solution could involve solving systems of ODEs or PDEs, depending on the modelling context, using either approximate or exact methods. We refer to this as the `basic method'.

In this article, we propose and test a modification to this method. Our modified method can be applied to models where the vector of parameters $\theta$ can be partitioned into $\theta=[\theta_\mathrm{outer}, \theta_\mathrm{inner}]$ such that there is a known relationship between the model solutions $y(\theta_\mathrm{outer}, \theta_\mathrm{inner})$ for parameters with the same value of $\theta_\mathrm{outer}$ but different values of $\theta_\mathrm{inner}$. This relationship can be expressed in the form of a transformation:
\begin{equation} \label{eq:scaling_relationship}
    y(\theta_\mathrm{outer}, \theta_\mathrm{inner}) = F\left( y(\theta_\mathrm{outer}, \theta_\mathrm{inner}^\mathrm{ref} ) {\blue ;  \theta_\mathrm{inner} } \right)  
\end{equation}
for some known function $F$, which we assume is relatively cheap to compute compared to solving the full forward model for $y(\theta)$. The simplest example is where the model solution is directly proportional to $\theta_\mathrm{inner}$, in which case this transformation is a simple linear scaling:
\begin{equation}
    y(\theta_\mathrm{outer},\theta_\mathrm{inner}) = \frac{\theta_\mathrm{inner}}{\theta_\mathrm{inner}^\mathrm{ref}}  y(\theta_\mathrm{outer}, \theta_\mathrm{inner}^\mathrm{ref}).  
\end{equation}

\begin{figure}
\centering 
    \includegraphics[width=\textwidth]{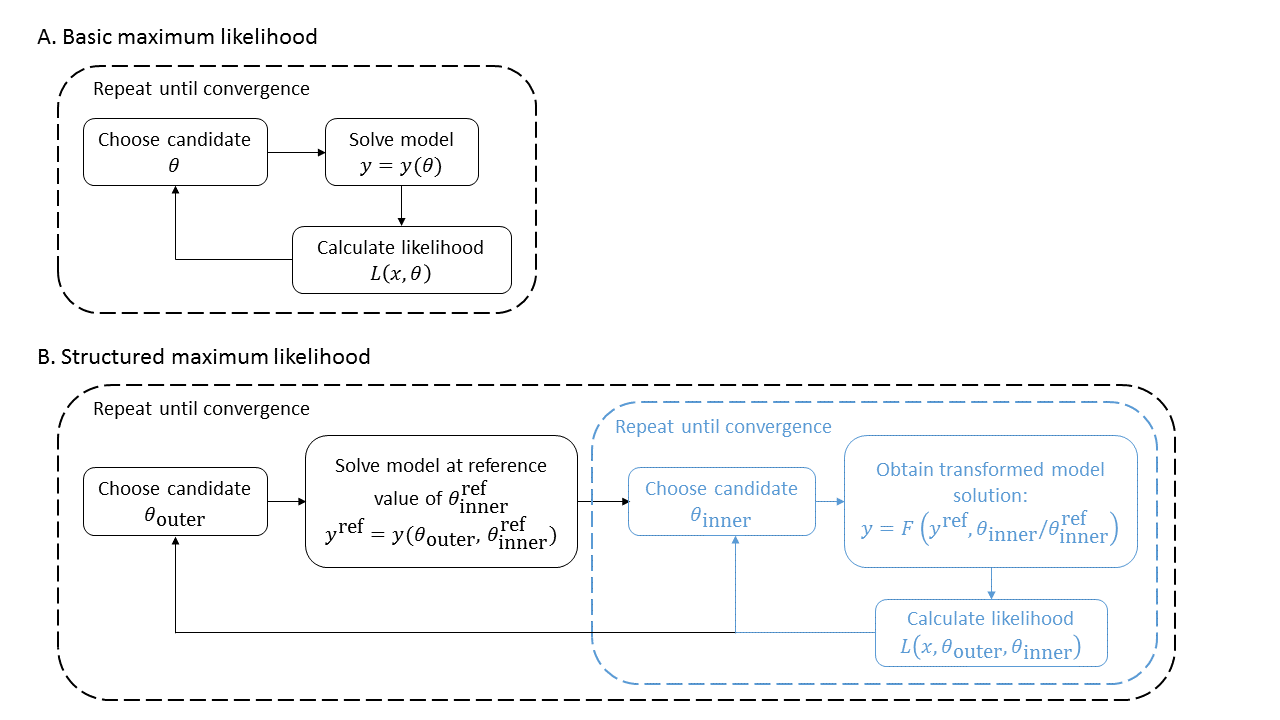}
    \caption{Schematic illustration of: (A) the basic maximum likelihood method and (B) the structured maximum likelihood method. The basic method is an $N$-dimensional optimisation problem for $\theta$. The structured method nests a $k$-dimensional inner optimisation problem for $\theta_\mathrm{inner}$ (blue) within an $(N-k)$-dimensional outer optimisation problem for $\theta_\mathrm{outer}$ (black). The inner optimisation problem can be solved numerically without rerunning the forward model to find $y$. Candidate values for parameters will typically be chosen using a standard optimisation algorithm.  }
    \label{fig:schematic}
\end{figure}

This type of scaling relationship enables a more efficient approach to maximum likelihood estimation because, for fixed $\theta_\mathrm{outer}$, the optimal value of $\theta_\mathrm{inner}$ may be calculated with only a single run of the forward model. Thus the $N$-dimensional optimisation problem in Eq. (\ref{eq:MLE_problem}) may be replaced by a nested pair of lower-dimensional problems:
\begin{eqnarray} 
\theta_{\mathrm{outer},\mathrm{MLE}} &=& \mathrm{argmax}_{\theta_\mathrm{outer}} L(x,\theta_\mathrm{outer}, \theta^*_\mathrm{inner} ), \label{eq:structured_problem1} \\
 \textrm{where } \qquad \theta^*_\mathrm{inner} &=& \mathrm{argmax}_{\theta_\mathrm{inner}} L(x,\theta_\mathrm{outer},\theta_\mathrm{inner}). \label{eq:structured_problem2}
\end{eqnarray}
To solve the inner optimisation problem in Eq. (\ref{eq:structured_problem2}), the full forward model only needs to be run once, for some defined reference value of $\theta_\mathrm{inner}=\theta_\mathrm{inner}^\mathrm{ref}$. The model solution $y^\mathrm{ref}= y(\theta_\mathrm{outer},\theta_\mathrm{inner}^\mathrm{ref})$ for this reference value of $\theta_\mathrm{inner}$ can then be transformed via Eq. (\ref{eq:scaling_relationship}) to find the model solution $y(\theta_\mathrm{outer},\theta_\mathrm{inner})$ for any value of $\theta_\mathrm{inner}$. We refer to this as the `structured method' (see Figure \ref{fig:schematic} for a schematic illustration of the basic and structured methods).

In this work, given a log-likelihood function, $L(x,\theta)$, we then re-scale to form a normalised log-likelihood $\hat{L}(x,\theta) =L(x,\theta) - L(x,\theta_{\textrm{MLE}})$, where $\theta_{\textrm{MLE}}$ is the MLE so that   $\hat{L}(x,\theta_{\textrm{MLE}})=0$.  Profile likelihoods are constructed by partitioning the full parameter $\theta$ into interest parameters $\psi$, and nuisance parameters $\omega$, such that $\theta = (\psi,\omega)$~\cite{pawitan2001}.  In this work we construct a series of univariate profiles by specifying the interest parameter to be a single parameter of interest in $\theta_\mathrm{outer}$, and the nuisance parameters are the remaining parameters in $\theta_\mathrm{outer}$.  For a fixed set of data $x$, the profile log-likelihood for the interest  parameter $\psi$ given the partition $(\psi,\omega)$ is
\begin{equation} \label{eq:profile_likelihood}
\hat{L}_p(\psi,x) = \sup_{\omega \mid \psi}\hat{L}(\psi, \omega, x),
\end{equation}
which implicitly defines a function $\omega^*(\psi)$ for optimal values of the nuisance parameters as a function of the target parameters for the given dataset. For univariate parameters this is simply a function of one variable.  For identifiable parameters these profile likelihoods will contain a single peak at the MLE. 

The degree of curvature of the log-likelihood function is related to inferential precision~\cite{pawitan2001}.  Since, in general, univariate profiles for identifiable parameters involve a single peak at the MLE, one way to quantify the degree of curvature of the log-likelihood function is to form asymptotic confidence intervals (CIs) by finding the interval where $\hat{L}_p \ge -\Delta_{q,n}/2$, where $\Delta_{q,n}$ denotes the $q$th quantile of the $\chi^2$ distribution with $n$ degrees of freedom~\cite{royston2007}. For univariate profiles, $n=1$. In this study we identify the interval in the interest parameter $\psi$ where $\hat{L}_p \ge -\Delta_{0.95,1}/2 = -1.92$, which identifies the 95\% CI.

\subsection*{Case studies}

\subsubsection*{Predator--prey model: Ecology}

Our first case study is a simple toy model for a prey species $R$ and predator species $P$, assuming a logistic growth for the prey and a type-II functional response (i.e. saturating at high prey density) for the predation term:
\begin{eqnarray}
\frac{dR}{dt} &=& rx \left(1-\frac{R}{K}\right) - \frac{aRP}{R+b},  \label{eq:prey} \\
\frac{dP}{dt} &=&  \frac{aRP}{R+b} - \mu P.                         \label{eq:pred} 
\end{eqnarray}
This is a type of Rosenzweig-MacArthur model \cite{rosenzweig1963graphical}, which is a generalisation of the classical Lotka-Volterra predator-prey model \cite{lotka1925elements,volterra1926variazioni}.

We suppose that some fixed fraction $p_\mathrm{obs}$ of the prey and the predator populations is observed on average at each time point $t$, and observed data $\hat{X}_t$ are Poisson distributed:
\begin{eqnarray}
    \hat{X}_{1,t} &\sim& \mathrm{Poisson}(y_{1,t} ), \\
    \hat{X}_{2,t} &\sim& \mathrm{Poisson}(y_{2,t} ), 
\end{eqnarray}
where $y_{1,t}=p_\mathrm{obs} R(t)$ and $y_{2,t}=p_\mathrm{obs} P(t)$ are the expected number of prey and predators observed respectively. For illustrative purposes, we assume the initial conditions and the parameters $K$ and $b$ are known (e.g. from information on the environmental conditions and average prey handling time), and perform inference on the parameter set $\theta=\left(r,a,\mu,p_\mathrm{obs}\right)$.

For the structured inference method, {\blue we use $p_\mathrm{obs}$ as the inner parameter  $\theta_\mathrm{inner}$.  We set the reference value of the inner parameter as $p_\mathrm{obs}^\mathrm{ref}=1$}, noting that if ${\bf y}_\mathrm{ref}$ is the model solution with $p_\mathrm{obs}=1$, then the solution for any given value of $p_\mathrm{obs}\in[0,1]$ is simply ${\bf y}=p_\mathrm{obs}{\bf y}_\mathrm{ref}$. 
Thus, under the structured method, the ordinary differential equation model (\ref{eq:prey})--(\ref{eq:pred}) only ever needs to be solved with  $p_\mathrm{obs}=1$. When evaluation of the model likelihood requires the expected value ${\bf y}$ of observed data for any other value of $p_\mathrm{obs}$, this is obtained via this simple linear scaling. {\blue Note that the choice to set $p_\mathrm{obs}^\mathrm{ref}=1$ is arbitrary and any fixed value of $p_\mathrm{obs}^\mathrm{ref}\in(0,1]$ could be chosen. }

\subsubsection*{SEIRS model: Epidemiology}

Our second case study is a SEIRS compartment model for an epidemic in a closed population of size $N$ \cite{diekmann2000mathematical}. We assume that the population at time $t$ is divided into compartments representing the number of people who are susceptible ($S$), exposed ($E$), infectious ($I$), and recovered ($R$). To model waning immunity, we subdivide the recovered compartment into two compartments $R_1$ and $R_2$ and assume that individuals transition from $R_1$ to $R_2$ and subsequently back to $S$. This is described by the following system of ordinary differential equations:
\begin{eqnarray}
\frac{dS}{dt} &=& -\lambda S + 2w R_2, \\
\frac{dE}{dt} &=& \lambda S - \gamma E, \\
\frac{dI}{dt} &=& \gamma E - \mu I, \\
\frac{dR_1}{dt} &=& \mu I - 2w R_1, \\
\frac{dR_2}{dt} &=& 2w (R_1-R_2), 
\end{eqnarray}
where $\gamma$, $\mu$ and $w$ are transition rates representing the reciprocal of the mean latent period, infectious period and immune period respectively, and $\lambda$ is the force of infection defined as
\begin{equation}
    \lambda(t) = \frac{R_0 \mu I(t)}{N}, 
\end{equation}
where $R_0$ is the basic reproduction number.
The model is initialised with a specified number of exposed individuals by setting $E(0)=E_0$, $S(0)=N-E_0$ and all other variables equal to zero. 

We assume that, on average, a fixed proportion $p_\mathrm{obs}$ of infections are observed. This could represent epidemiological surveillance data, for example the number of notified cases under the assumption that surveillance effort is steady over time. Alternatively it could represent a particular clinical outcome, such as hospital admission. Unknown case ascertainment or uncertain clinical severity are common, particularly for new or emerging infectious diseases. Hence, joint inference of this parameter with parameters of the SEIRS model, in situations where these are identifiable, is an important task in outbreak modelling \cite{lustig2023modelling,russell2020reconstructing,flaxman2020estimating}.

Observations are not typically recorded instantaneously at the time of infection, so we assume there is an average time lag of $1/\alpha$ from the time an individual becomes infectious to the time they are observed. These assumptions are modelled via the differential equations
\begin{eqnarray}
    \frac{dC_1}{dt} &=& p_\mathrm{obs} \gamma E - \alpha C_1, \\
    \frac{dC_2}{dt} &=& \alpha C_1.
\end{eqnarray}
Here, $C_1$ represents the number of infected individuals who will be observed but have not yet been observed, and $C_2$ is the cumulative number of observed infections. The expected number of observations per day at time $t$ is denoted $y_t$ and is equal to $\alpha C_1$.

We assume that the number of observed cases $\hat{X}_t$ on day $t$ is drawn from a negative binomial distribution {\blue (which is a commonly used distribution in epidemic modelling when observed cases may have more stochasticity than under a binomial or Poisson distribution \cite{abbott2020estimating,golding2023modelling})} with mean $y_t$ and dispersion factor $k$:
\[
\hat{X_t} \sim \textrm{NegBin}(y_t, k).
\]
We assume that the initial conditions and the transition rates $\gamma$, $\mu$ and $\alpha$ are known (e.g. from independent epidemiological data on the average latent period and infectious period), and perform inference on the parameter set $\theta=\left( R_0, w, p_\mathrm{obs},k\right)$. 

Similar to the predator--prey model, {\blue we use $p_\mathrm{obs}$ as the inner parameter $\theta_\mathrm{inner}$ and set $p_\mathrm{obs}^\mathrm{ref}=1$}, noting that if ${\bf y}_\mathrm{ref}$ is the model solution with $p_\mathrm{obs}=1$, then the solution for any given value of $p_\mathrm{obs}\in[0,1]$ is simply ${\bf y}=p_\mathrm{obs}{\bf y}_\mathrm{ref}$. {\blue Again, the choice to set $p_\mathrm{obs}^\mathrm{ref}=1$ is arbitrary and any fixed value of $p_\mathrm{obs}^\mathrm{ref}\in(0,1]$ could be chosen. }

\subsubsection*{Advection-diffusion-reaction model: Environmental science}

In the context of environmental modelling and pollution management, advection-diffusion-reaction models are often used to study how dissolved solutes are spatially distributed over time.  Very often the equation for the solute concentration $u(\mathbf{x},t)$ at position $\mathbf{x}$ and time $t$ includes a source term to describe chemical reactions~\cite{Herzer1989}.  A common approach is to assume that chemical reactions are fast relative to transport and a linear isotherm model.  This gives one of the most common models of reaction-transport.
\begin{equation} \label{eq:adv_diff_pde}
\dfrac{\partial u}{\partial t} = D  \nabla^2 u -v \cdot \nabla u - \dfrac{\partial s}{\partial t},
\end{equation}
where $D>0$ is the diffusivity, often called the dispersion coefficient in solute transport modelling literature, $v$ is the advection velocity, and $\partial s / \partial t$ is a sink term that represents a chemical reaction that converts $u(x,t)$ into some immobile product $s(x,t)$. A common form for the sink term is $\partial s / \partial t = f u - g s$, where $f>0$ is the forward reaction rate and $g>0$ is the backward reaction rate. In the limiting case where the chemical reactions occur instantaneously, we have $s = fu/g$ which we rewrite as $s= (R-1)u$ where $R=1+f/g>1$ is a dimensionless parameter known as the retardation factor. Substituting this relationship into Eq. (\ref{eq:adv_diff_pde}) gives
\begin{equation}
R\dfrac{\partial u}{\partial t} = D \nabla^2 u -v\cdot\nabla u.
\end{equation}
This shows that the presence of chemical reactions effectively retards the dispersion and advection processes since the chemical reaction leads to effective diffusion and advection rates of $D^*=D/R$ and $v^*=v/R$, respectively.  

While, in general, this transport equation can be solved numerically for a range of initial and boundary conditions, exact solutions can also be used where appropriate.  In one spatial dimension with initial condition $u(x,0) = 0$ (for $0 \le x < \infty$) and boundary condition $u(0,t)=U_0$, the exact solution is~\cite{Ogata1961,Genuchten1976}
\begin{eqnarray}
 u(x,t) & =  & \dfrac{U_0}{2}\left[\textrm{erfc}\left(\dfrac{x-vt/R}{\sqrt{4Dt/R}}\right)+ \textrm{exp}\left(\dfrac{vx }{D}\right)\textrm{erfc}\left(\dfrac{x+vt/R}{\sqrt{4Dt/R}}\right)  \right], \label{eq:ogata_banks1} \\
s(x,t) &=& (R-1)u(x,t). \label{eq:ogata_banks2}
\end{eqnarray}

We assume that the solute species $u$ is not observed directly, and observations of the product species $s$ are taken at a fixed time $t_\mathrm{obs}$ and at a series of points in space $x=x_i$, and are subject to Gaussian noise with standard deviation $\sigma$. Thus, observed data $\hat{X}$ take the form:
\begin{equation}
\hat{X}_i \sim N\left( s(x_i,t_\mathrm{obs}), \sigma\right). 
\end{equation}
We assume that the solute boundary concentration $U_0$ is known and perform inference on the parameter set $\theta = \left(D, v, R, \sigma\right)$.  In this instance we take the most fundamental approach in forming a likelihood function by assuming that observations are normally distributed about the solution of the PDE model with a constant standard deviation.  This is a commonly--employed approach, however our likelihood-based approach can be employed using a range of measurement models~\cite{murphy2024}.

For the structured method, we exploit a symmetry in the analytical solution in Eq.(\ref{eq:ogata_banks1}), namely that
\begin{equation} \label{eq:ogata_banks_transform}
u(x,t;R) = u\left(x, (R^\mathrm{ref}/R) t  ; R^\mathrm{ref} \right).
\end{equation}
{\blue We therefore use $R$ as the inner parameter $\theta_\mathrm{inner}$. We set $R^\mathrm{ref}=1$} and note that if $u^\mathrm{ref}$ is the model solution for $u(x,t)$ when $R=R^\mathrm{ref}=1$ then, by Eqs. (\ref{eq:ogata_banks2}) and (\ref{eq:ogata_banks_transform}), the solution for the product $s(x,t)$ for an arbitrary value of $R>1$ is
\begin{equation} \label{eq:ogata_banks_transformed_product}
s(x,t)= (R-1) u^\mathrm{ref}(x, t/R).
\end{equation}
This relationship between the inner parameter $R$ and the model solution is more complicated than the simple multiplicative scaling seen in the first two case studies. As Eq. (\ref{eq:ogata_banks_transform}) shows, changing the value of $R$ is equivalent to rescaling one of the independent variables ($t$) in the PDE for $u(x,t)$. Therefore, evaluating the solution $s(x,t)$ for a different value of $R$ is equivalent to evaluating the reference solution $u^\mathrm{ref}(x,t)$ at a different value of $t$, as well as changing the multiplicative factor that relates $u(x,t)$ and $s(x,t)$.  Since a numerical scheme for an advection--diffusion PDE will typically generate solution values at a series of closely spaced time points, it will generally be straightforward to query the reference solution $u^\mathrm{ref}$ at an earlier time point. This allows the solution for arbitrary values of $R>1$ to be obtained without costly re-evaluation of the numerical solution scheme. 
 
To emulate this situation, for given values of the outer parameters $D$ and $v$, we 
generate the reference solution $u^\mathrm{ref}$ by evaluating Eq. (\ref{eq:ogata_banks1}) with $R=1$ at a series of equally spaced time points $t=t_i$ from $t=0$ to $t=t_\mathrm{obs}$. When the solution for $s(x,t)$ in Eq. (\ref{eq:ogata_banks_transformed_product}) subsequently requires $u^\mathrm{ref}$ to be queried at $t=t_\mathrm{obs}/R$, we approximate this using linear interpolation between the nearest two values of $t_i$. {\blue Here, choosing $R^\mathrm{ref}=1$ is beneficial because it means that the reference solution $u^\mathrm{ref}$ only ever needs to be queried at values of $t$ within the range of the numerical solution, $t\in[0, t_\mathrm{obs}]$. }

We note that, although we have studied the one-dimensional version of this model for illustrative purposes, our approach could also be used in higher-dimensions. Under these circumstances, numerically solving the PDE is more computationally expensive, so any efficiency improvement from taking advantage of the model structure would be be amplified.

\subsection*{Computational methods}
{\blue For each model case study, we first solved the underlying deterministic model equations with fixed parameter values (see Supplementary Table S1). This was done via numerical solution of the ODEs for the first two case studies, whereas the analytical solution was used for the third case study. Once the deterministic solution $y$ was calculated, we then generated a synthetic dataset $x$ by simulating the specified noise model. }

To perform parameter inference using the basic method, we first found the maximum likelihood estimate (MLE) for the target parameter set by solving the optimisation problem in Eq. (\ref{eq:MLE_problem}). We then computed univariate likelihood profiles for each target parameter $\psi$ by taking a uniform mesh over some interval $[\psi_{\textrm{min}}, \psi_{\textrm{max}}]$ with a modest number of mesh points ($n_\mathrm{mesh}=41$) and solving Eq. (\ref{eq:profile_likelihood}) to find the profile log-likelihood at each mesh point. In practice, we determined an appropriate interval $[\psi_{\textrm{min}}, \psi_{\textrm{max}}]$ by computing $\hat{L}_p$ across a trial interval and if necessary widening the interval until it contained the 95\% CI. 

For the first mesh point to the right of the MLE, the optimisation algorithm was initialised using the MLE values for each parameter. For each subsequent mesh point stepping from left to right, the algorithm was initialised using the previous profile solution. This procedure was then repeated for mesh points to the left of the MLE by stepping right to left. The boundaries of the 95\% CI were estimated using one-dimensional linear interpolation to compute the value of the target parameter at which the normalised profile log-likelihood equaled the threshold value $-\Delta_{0.95,1}/2$.

We used a similar procedure for the structured inference method, finding the MLE by solving the nested optimisation problem in Eq. (\ref{eq:structured_problem1})--(\ref{eq:structured_problem2}) (see Figure \ref{fig:schematic}B). We then constructed univariate profiles as described above, but with Eq. (\ref{eq:profile_likelihood}) similarly recast as a nested pair of optimisation problems. This means that, with the structured method, there is one fewer parameter to profile because for a given dataset $x$, the inner parameter is treated as a deterministic function of the outer parameters and so does not need to be profiled separately. 

All optimisation problems were solved numerically using the constrained optimisation function {\em fmincon} in Matlab R2022b with the interior point algorithm and default tolerances. {\blue We calculated the relative error between the MLE and the true parameter values, and the number of calls to the forward model required by both the basic and structured methods. This enabled us to compare the accuracy and efficiency of the two methods. Because the results will depend on the particular realisation of the noise model, we repeated this process $M=500$ times for each case study, calculating the MLE and 95\% CIs for $M$ independent datasets using fixed parameter values (see Supplementary Table S1). We report the median and interquartile range of the relative error and the number of function calls across the $M$ datasets. We also report the proportion of the $M$ datasets for which the 95\% CI for each parameter contained the true value.} We also carried out sensitivity analysis by randomly varying the model parameters for each synthetically generated dataset (see Supplementary Material for details).

Documented Matlab code that implements both the basic and the structured method on the three models described above is publicly available at: 
\url{https://github.com/michaelplanknz/structured-inference}. The code can also be run on a user-supplied model, either with synthetically generated data from the specified model or with user-supplied data. This requires the user to provide a function that returns the model solution for specified parameter values, and a separate function that transforms the reference model solution to the model solution for a specified value of the inner parameters. The user can choose from a range of pre-supplied noise models, including additive or multiplicative Gaussian, Poisson, and negative binomial. Detailed instructions for running the method on a user-supplied models and/or data are available in the ReadMe file at the URL above.

\section*{Results}

\begin{figure}
    \centering
    \includegraphics[width=\textwidth]{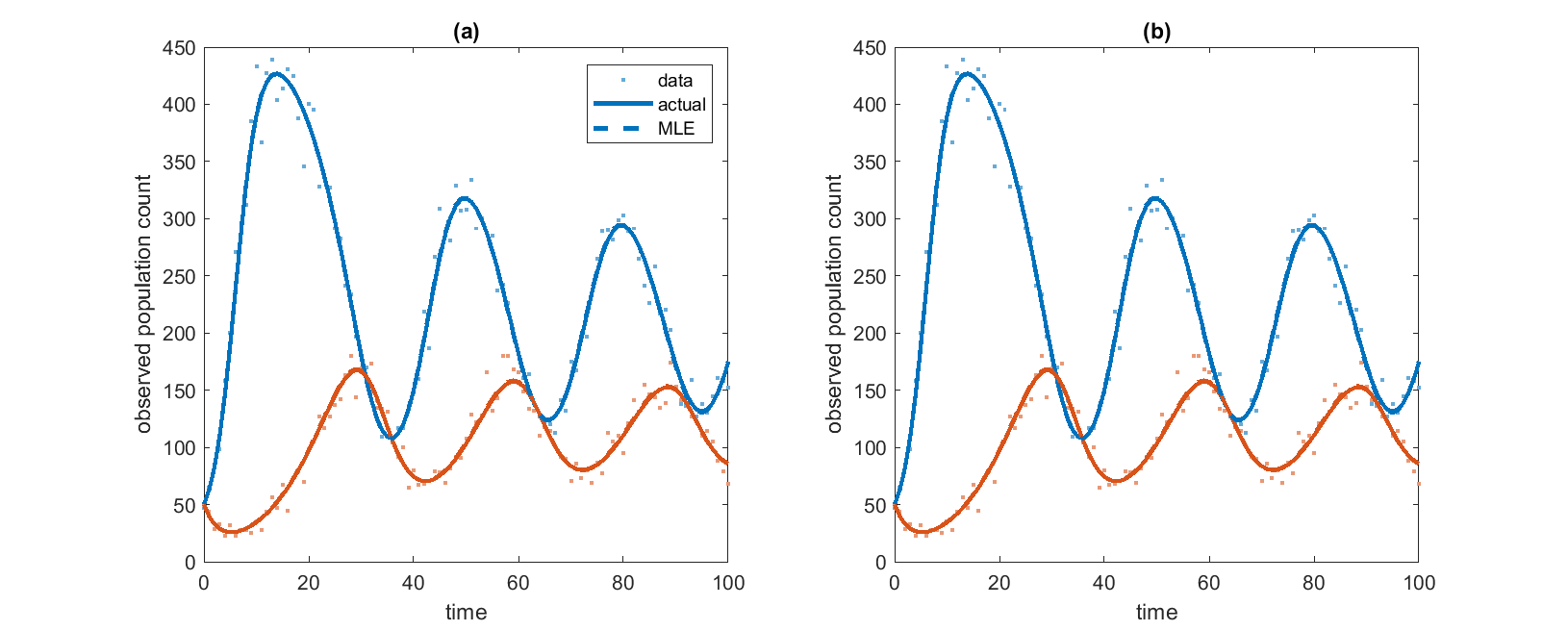}
    \caption{ Results for the predator--prey model from: (a) the basic method; (b) the structured method. Each panel shows the solution under the actual parameter values (solid curves); the solution under the maximum likelihood estimate for the parameter values (dashed curves); and the simulated data (dots) for the prey (blue) and predator (red). Where the dashed curve representing the solution at the MLE is not visible, this is because it coincides with the solid curve representing the solution at the true parameter values. }
    \label{fig:MLE_LV}
\end{figure}

\begin{figure}
    \centering
    \includegraphics[width=\textwidth]{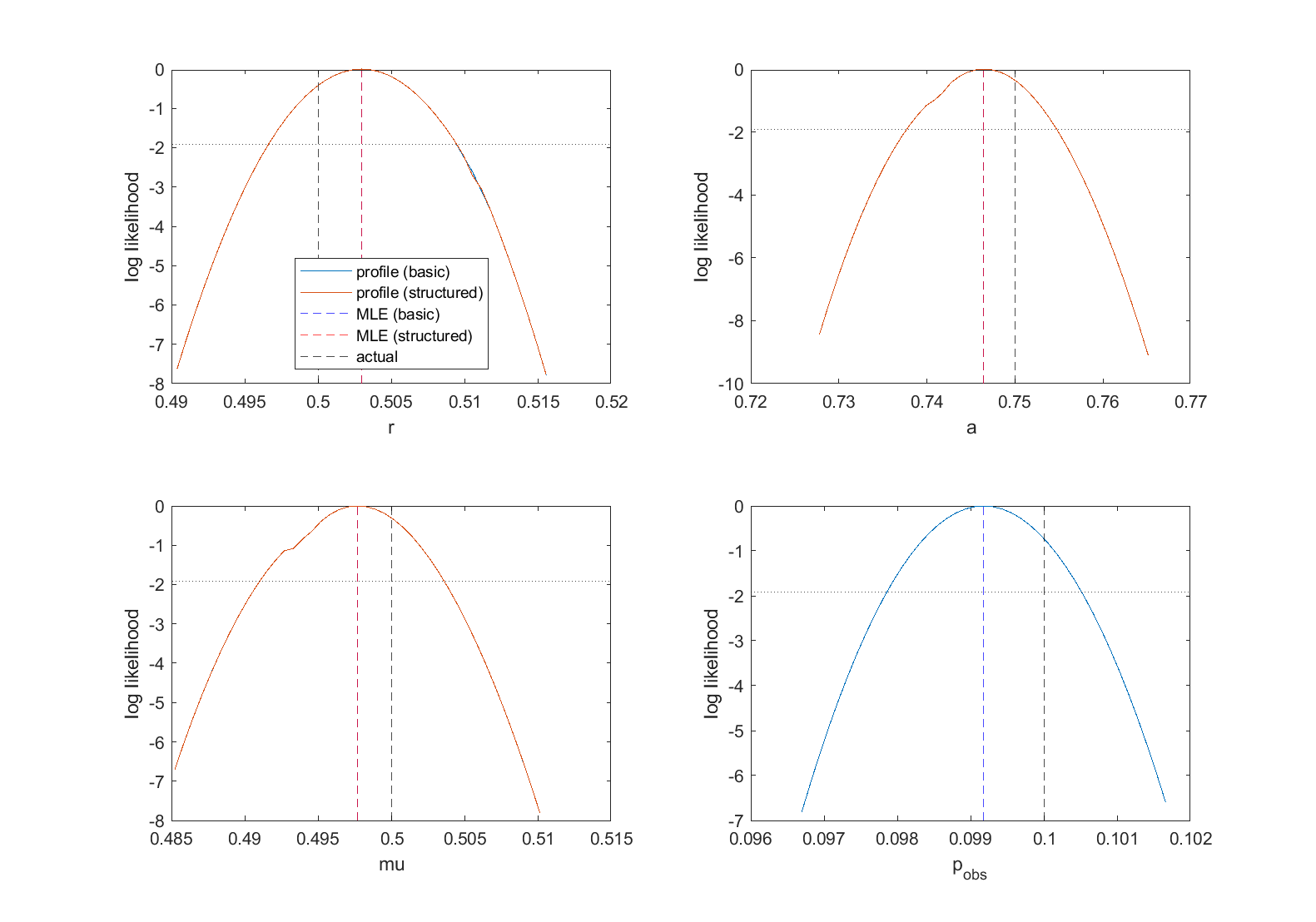}
    \caption{Normalised likelihood profiles for inferred parameters of the predator--prey model: prey intrinsic growth rate ($r$); predation coefficient ($a$); predator death rate ($\mu$); observation probability ($p_\mathrm{obs}$). The blue curves are  from the basic method; the red curves are from the structured method; dashed vertical lines indicate the actual values (black) and the maximum likelihood estimates for each parameter under the basic method (blue) and structured method (red). Where the blue curve is not visible, this is because it coincides exactly with the red curve. Dotted horizontal line shows the threshold normalised log likelihood for the 95\% confidence interval. }
    \label{fig:profiles_LV}
\end{figure}

\subsubsection*{Predator-prey model}
The model exhibited limit cycle dynamics for the parameter values chosen, completing around three cycles during the time period observed (Figure \ref{fig:MLE_LV}). Both the basic and the structured method were able to successfully recover the correct parameter values from the observed data. The univariate likelihood profiles were almost identical for both methods and were unimodal, implying that the parameters are identifiable (Figure \ref{fig:profiles_LV}). The 95\% confidence intervals (range of values above the dotted horizontal line in Figure \ref{fig:profiles_LV}) contained the true parameter value for all target parameters for both methods. Obtaining the same results from both methods was an expected result, which provided confirmation that the structured method was working correctly.

When run on $M=500$ independently generated synthetic datasets, both methods had similar levels of accuracy in the MLE, with a median relative error of $0.5$\% (interquartile range [0.3\%, 0.7\%]) -- see Table \ref{tab:results}. The coverage properties of both methods were good, with the 95\% CI containing the true parameter value in 94--96\% of cases for both methods. However, the structured method required only 7492 calls (i.e. evaluations of the forward model) on average, compared to 15054 for the basic method. The reduction in the number of calls for the structured method compared to the basic method was 50.2\% (interquartile range [48.5\%, 52.1\%]). The efficiency improvement was greater in the profile likelihood component of the algorithm than in calculation of the MLE, where the reduction in calls was 18.4\% [6.3\%, 28.8\%]). This is partly because, with the structured method, the inner parameter is effectively a function of the outer parameters (i.e. the solution of the inner optimisation problem in Eq. (\ref{eq:structured_problem2})), so does not need to be profiled separately. However, there was also a reduction in the number of calls per parameter profile.

\begin{table}
\linespread{1.0}
\small
\centering
\begin{tabular}{lllll} 
\hline  
& & \multicolumn{2}{l}{\bf Relative error (\%)} & \\ 
 & & Basic & Structured & \\ 
Predator-prey && 0.5 [0.3, 0.7] & 0.5 [0.3, 0.7] & \\ 
SEIRS && 2.6 [1.3, 4.9] & 2.6 [1.2, 4.6] & \\ 
Adv. diff. && 11.0 [6.6, 17.3] & 11.0 [6.6, 17.3] & \\ 
\hline  
& & \multicolumn{2}{l}{\bf 95\% CI coverage} & \\ 
  && Basic & Structured \\ 
Predator-prey & $r$ & 94.8\% & 94.8\%  \\ 
& $a$ & 94.6\% & 94.0\%  \\ 
& $\mu$ & 95.4\% & 95.2\%  \\ 
& $p_{obs}$ & 95.4\% & - \\ 
SEIRS & $R_0$ & 89.0\% & 92.2\%  \\ 
& $w$ & 87.2\% & 90.4\%  \\ 
& $p_{obs}$ & 86.2\% & - \\ 
& $k$ & 88.2\% & 91.8\%  \\ 
Adv. diff. & $D$ & 89.8\% & 89.8\%  \\ 
& $v$ & 90.8\% & 90.8\%  \\ 
& $R$ & 90.2\% & - \\ 
& $\sigma$ & 88.4\% & 88.4\%  \\ 
\hline  
&& \multicolumn{3}{l}{\bf Function calls (MLE) } \\ 
  && Basic & Structured & Improvement (\%) \\ 
Predator-prey && 208 [194, 228] & 171 [153, 194]  & 18.4 [6.3, 28.8] \\ 
SEIRS && 143 [134, 154] & 130 [121, 143]  & 9.2 [-2.6, 19.5] \\ 
Adv. diff. && 173 [158, 191] & 98 [88, 111]  & 42.9 [33.9, 50.3] \\ 
\hline  
&& \multicolumn{3}{l}{\bf Function calls (profiles) } \\ 
  && Basic & Structured & Improvement (\%) \\ 
Predator-prey && 14839 [14084, 15815] & 7320 [7017, 7662]  & 50.7 [49.0, 52.7] \\ 
SEIRS && 13389 [12854, 13750] & 7339 [7123, 7545]  & 44.9 [42.6, 46.7] \\ 
Adv. diff. && 10247 [9699, 10893] & 4675 [4339, 5059]  & 54.6 [51.5, 56.9] \\ 
\hline  
&& \multicolumn{3}{l}{\bf Function calls (total) } \\ 
  && Basic & Structured & Improvement (\%) \\ 
Predator-prey && 15054 [14294, 16029] & 7492 [7184, 7834]  & 50.2 [48.5, 52.1] \\ 
SEIRS && 13542 [12998, 13900] & 7476 [7251, 7693]  & 44.5 [42.3, 46.3] \\ 
Adv. diff. && 10424 [9870, 11064] & 4772 [4433, 5158]  & 54.4 [51.2, 56.7] \\ 
\hline  
\end{tabular}  
\caption{Relative error in the maximum likelihood estimate, coverage of the 95\% CI (i.e. proportion of datasets for which the 95\% CI contained true parameter value) for each profiled parameter, and the number of calls to the forward model solver (number to calculate the MLE, number to calculate the likelihood profiles, and total number) under the basic and structured inference methods for each of the three model case studies. Note coverage statistics for the inner parameters ($p_\mathrm{obs}$ and $R$) are not applicable for the structured method, as the inner parameter is calculated as a function of the outer parameters. The `Improvement' column shows the reduction in the number of function calls under the structured method relative to the basic method. Results show the median and interquartile range across $M=500$ independently generated datasets for each model.}
\label{tab:results}
\end{table}

\begin{figure}
    \centering
    \includegraphics[width=\textwidth]{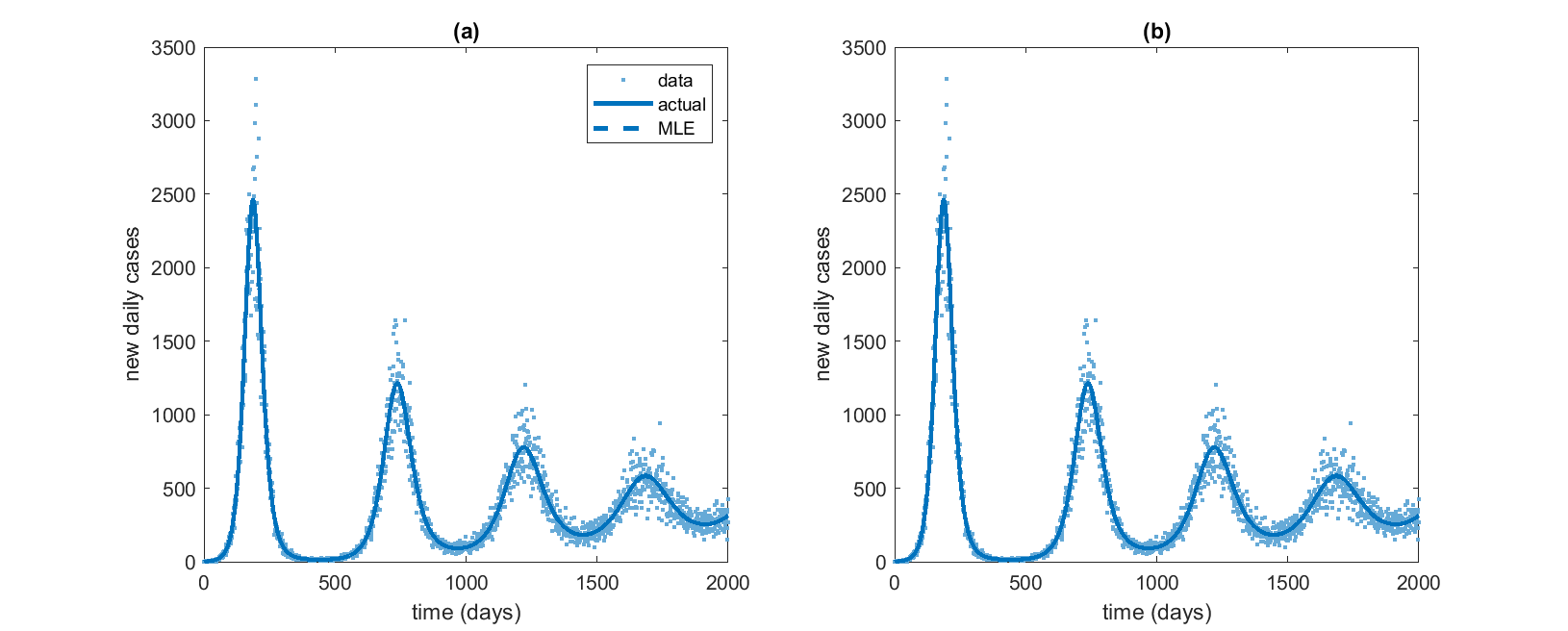}
    \caption{ Results for the SEIRS model from: (a) the basic method; (b) the structured method. Each panel shows the solution under the actual parameter values (solid curves); the solution under the maximum likelihood estimate for the parameter values (dashed curves); and the simulated data (dots). Where the dashed curve representing the solution at the MLE is not visible, this is because it coincides with the solid curve representing the solution at the true parameter values.}
    \label{fig:MLE_SEIR}
\end{figure}

\begin{figure}
    \centering
    \includegraphics[width=\textwidth]{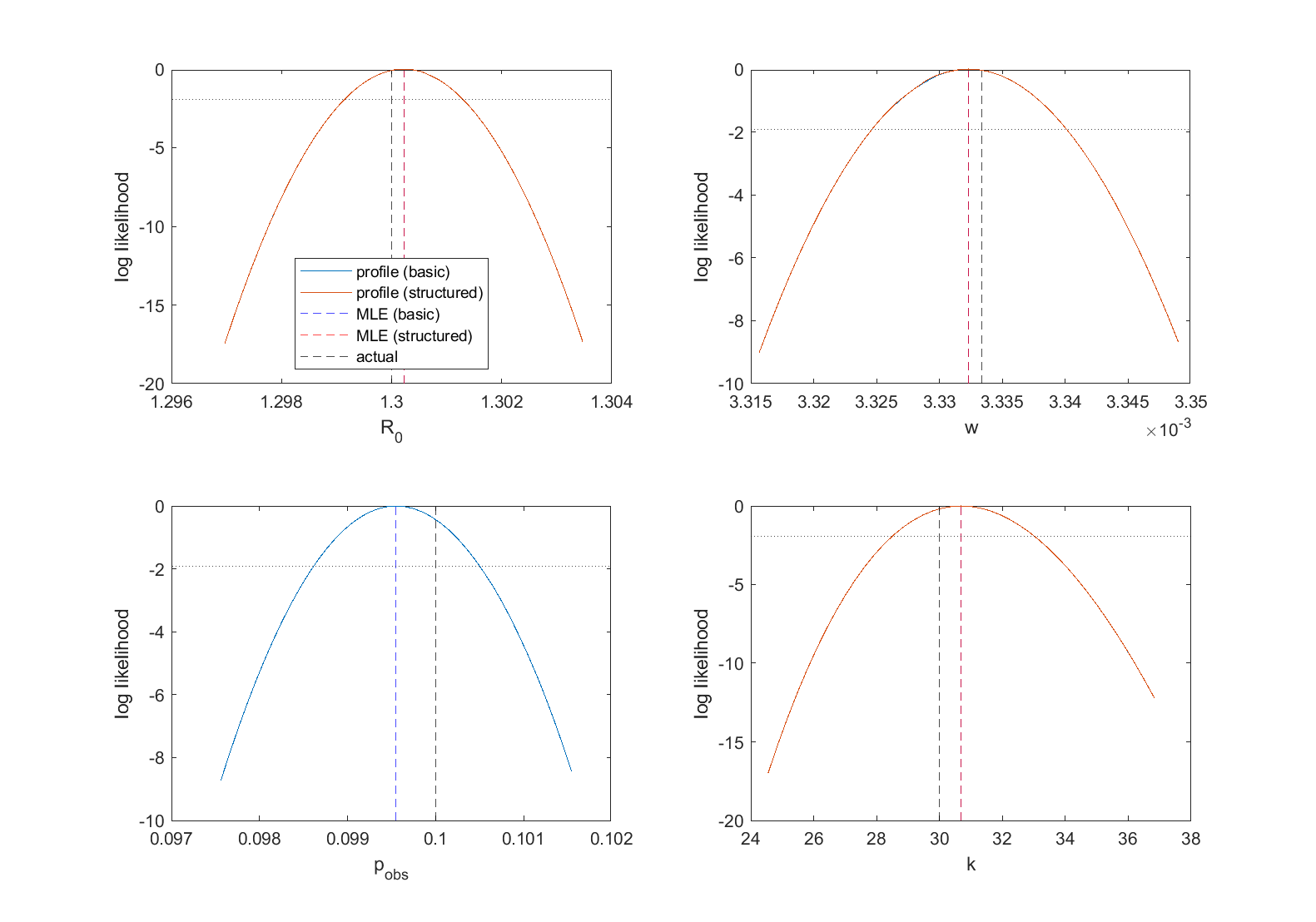}
    \caption{Normalised likelihood profiles for inferred parameters of the SEIRS model: basic reproduction number ($R_0$); recovered to susceptible rate ($w$); observation probability ($p_\mathrm{obs}$); negative binomial dispersion parameter for observed data ($k$). The blue curves are  from the basic method; the red curves are from the structured method; dashed vertical lines indicate the actual values (black) and the maximum likelihood estimates for each parameter under the basic method (blue) and structured method (red). Where the blue curve is not visible, this is because it coincides exactly with the red curve. Dotted horizontal line shows the threshold normalised log likelihood for the 95\% confidence interval.}
    \label{fig:profiles_SEIR}
\end{figure}

\subsubsection*{SEIRS model}
The model exhibits a large epidemic wave, followed by a series of successively smaller waves as the model approaches the stable endemic equilibrium (Figure \ref{fig:MLE_SEIR}). Again, both the basic and the structured method successfully recovered the correct parameter values, and produced almost identical MLEs and univariate likelihood profiles (Figure \ref{fig:profiles_SEIR}). The profiles were unimodal, meaning that the parameters were identifiable.

When run on $M=500$ independently generated datasets, both methods again had similar accuracy, with a relative error in the MLE of 2.6\% [1.3\%, 4.9\%] for the basic method, compared to 2.6\% [1.2\%, 4.6\%] for the structured method (Table \ref{tab:results}). The coverage rates were slightly better for the structured method (90--93\%) compared to the basic method (86--90\%). However, the main advantage of the structured method was its efficiency, with a 44.5\% [42.3\%, 46.3\%] reduction in the total number of calls required compared to the basic method. As for the predator-prey model, most of the efficiency improvement was in the profile likelihood component, although there was also a reduction of 9.2\% [-2.6\%, 19.5\%] in the number of calls required for the MLE.

\begin{figure}
    \centering
    \includegraphics[width=\textwidth]{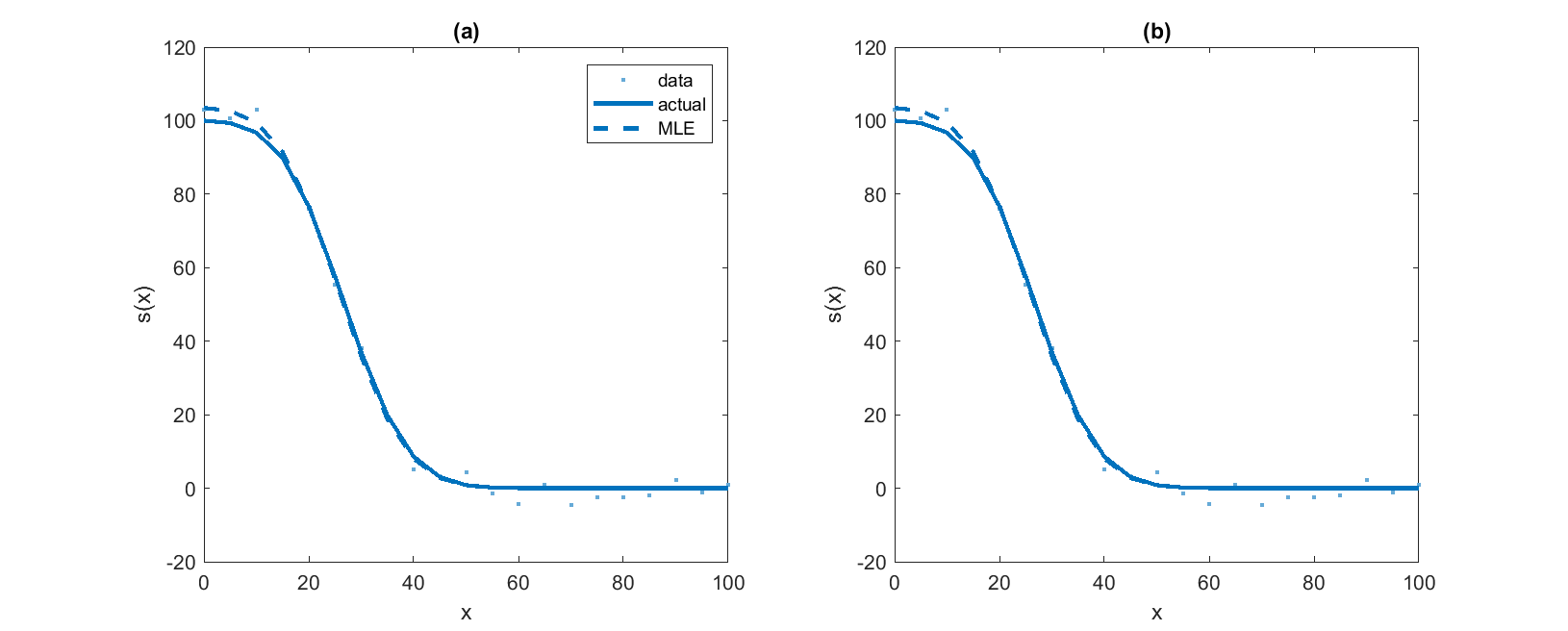}
    \caption{ Results for the advection--diffusion model from: (a) the basic method; (b) the structured method. Each panel shows the solution under the actual parameter values (solid curves); the solution under the maximum likelihood estimate for the parameter values (dashed curves); and the simulated data (dots). Where the dashed curve representing the solution at the MLE is not visible, this is because it coincides with the solid curve representing the solution at the true parameter values.}
    \label{fig:MLE_PDE}
\end{figure}

\begin{figure}
    \centering
    \includegraphics[width=\textwidth]{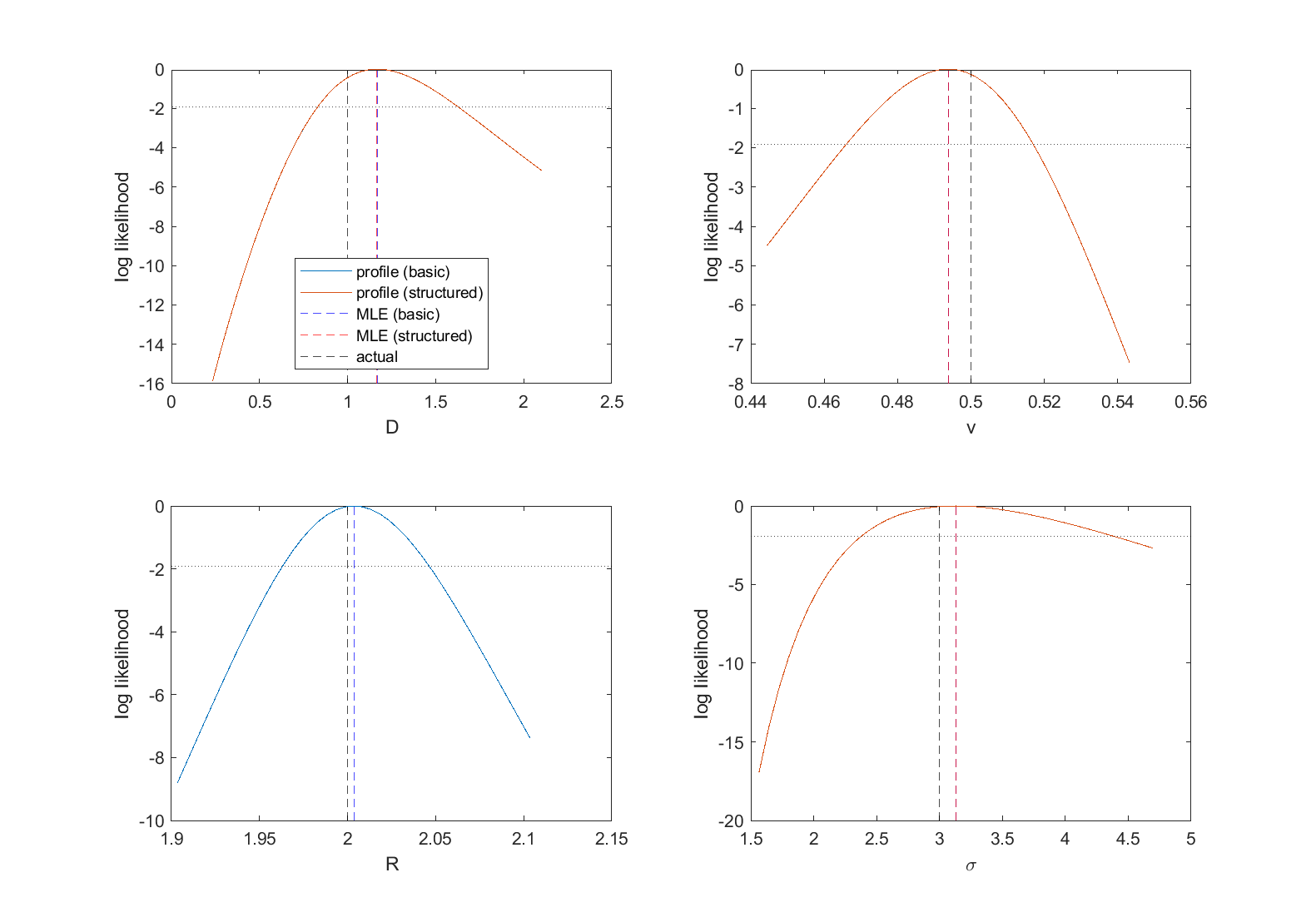}
    \caption{Normalised likelihood profiles for inferred parameters of the advection--diffusion model: diffusion coefficient ($D$); advection velocity ($v$); retardation factor ($R$); standard deviation of noise in observed data ($\sigma$). The blue curves are  from the basic method; the red curves are from the structured method; dashed vertical lines indicate the actual values (black) and the maximum likelihood estimates for each parameter under the basic method (blue) and structured method (red). Where the blue curve is invisible, this is because it coincides exactly with the red curve. Dotted horizontal line shows the threshold normalised log likelihood for the 95\% confidence interval.}
    \label{fig:profiles_PDE}
\end{figure}

\subsubsection*{Advection-diffusion model}
The model exhibits a sigmoidal decay in concentration with distance from the source at the observation time of $t_\mathrm{obs}=100$. (Figure \ref{fig:MLE_PDE}). Again, both methods successfully recovered the correct parameter values. Although all univariate likelihood profile were unimodal, the confidence intervals for the diffusion coefficient $D$ and noise magnitude $\sigma$ (range of values above the dotted horizontal line in Figure \ref{fig:profiles_PDE}) were relatively wide compared to the other target parameters (Figure \ref{fig:profiles_PDE}). This indicates that $D$ and $\sigma$ were relatively weakly identified by the data, and a range of values of these parameters were approximately consistent with the observed data.

When run on $M=500$ independently generated datasets, the relative error was 11.0\% [6.6\%, 17.3\%] for both methods (Table \ref{tab:results}). Coverage rates were reasonable (88--91\% for both methods). The structured method required 54.4\% [51.2\%, 56.7\%] fewer calls to the model than the basic method did. Even for the MLE component alone, the structured method provided a substantial improvement in efficiency with 42.9\% [33.9\%, 50.3\%] fewer calls than the basic method.

\subsubsection*{Sensitivity analysis}
The results described above for each model are for $M=500$ independent datasets, each generated using the same set of parameter values. To test the sensitivity of our results to the true parameter values, we repeated the analysis with $M=500$ datasets, each generated from an independently chosen set of parameter values. For each target parameter, we chose the value from an independent normal distribution with the same mean as previously (see Supplementary Table S1) and with coefficient of variation $0.1$. 

The results are similar to those described above for fixed parameters. The accuracy and coverage properties are similar for both methods, but the structured method is consistently more efficient than the basic method (see Supplementary Material for details).

\section*{Discussion}
We have developed an adapted method for likelihood-based parameter inference and uncertainty quantification. The method is based on standard maximum likelihood estimation and profile likelihood, but exploits known structure in the mechanistic model to split a high-dimensional optimisation problem into a nested pair of lower-dimensional problems using a similar approach to \cite{loos2018hierarchical}. Only the outer problem requires the forward model to be solved at each step. Like standard maximum likelihood estimation, our method can be used with a range of different types of mathematical models and noise models, provided a likelihood function exists.

We have illustrated the new method on three toy models from different areas of the life sciences: a predator-prey model; a compartment-based epidemic model; and an advection-diffusion PDE model. We have shown that the standard method and the new structured method provided comparable levels of accuracy in parameter estimates and coverage of inferred confidence intervals. However, the structured method was consistently more efficient, requiring substantially fewer calls to the forward model. In more complex models that are computationally expensive to solve, this would provide a major improvement in computation time. 

Although we have illustrated our method on three simple models, each with a single inner parameter, we anticipate there will be a broad class of models for which our approach could be applied. These include models where one or more dependent variables are linear in one or more parameters, as considered by \cite{loos2018hierarchical,schmiester2020efficient}. This was the situation in the first two of our case studies. A similar situation would arise if observed model components were linear in an initial or boundary condition, or in a model with one-way coupling from a nonlinear to a linear component. 
Our approach is also applicable to models where a parameter effectively rescales one of the independent variables (typically space or time), which was the situation in the third case study. 
More generally, models which possess some sort of symmetry or invariance may also be candidates for applying our approach. These include travelling wave solutions (such as in models of a spreading population or other reaction--diffusion equations \cite{baker2007travelling,baker2012models}),  similarity solutions (such as in models of chemotaxis \cite{rascle1995finite,byrne1998effect}), or scale invariance (such as in size-spectrum models of marine ecosystem dynamics \cite{capitan2010scale,plank2012ecological}). Exploring the range of models and classes of parameter relationships for which our methodology is applicable is an avenue for further research.

The process of specifying inner parameters and identifying the transformation relationship will in general be model-dependent. In models where such relationships exist, they may be revealed by a non-dimensionalisation procedure, which is a well-established technique in mathematical modelling \cite{murray2003mathematical}.  A clear example of this would include a broad class of reaction--diffusion models with a logistic source term with carrying capacity $K$, where changing the value of $K$ linearly rescales the solution of the model~\cite{Haridas2017}. {\blue However, we acknowledge that not all models will be amenable to the structured inference we have presented because the relationship between parameters of interest and model solution will be unknown or too complex to identify {\em a priori}. }

Our new method does not avoid the risk of the optimisation routine converging to a local minimum instead of a global minimum. Nonetheless, one way to tackle this is to use a global search algorithm, and using our approach to reduce the dimensionality of the parameter space may help this to succeed. 

We have examined case studies in which data come from some specified noise process applied to the solution of an underlying deterministic model. {\blue In the second and third case studies, where the noise distribution had a variance parameter, we did not assume this was known {\em a priori}, but included it as a target for inference for both the basic and the structured method.} Profile likelihood has also been applied to stochastic models, for example a stochastic model of diffusion in heterogeneous media \cite{simpson2021profile}. Another feature of our study is that we have applied our methods to standard cases where we treat the available data as fixed, but we note that we could adapt our method to deal with cases where inference and identifiability analysis is dynamically updated as new data becomes available~\cite{Cassidy2023}. How the size of the efficiency improvement scales with the number of inner and outer parameters in more complex models is also an interesting question for future work.

In this article, we have focused on parameter estimation and identifiability analysis. Given our univariate profile likelihood functions, it is possible to make profile-wise predictions, propagating uncertainty in parameter estimates through to uncertainty in model predictions~\cite{simpson2023profile}. This enables understanding of how variability in different parameters impacts the solution of the mathematical model. Given that our structured profile likelihood functions can be computed with far less computational overhead than the standard approach, it is possible to use our structured approach to speed up the calculation of likelihood-based prediction intervals~\cite{simpson2023profile}.

All results in this study focus on assessing practical identifiability since we are interested in real-world problems, which involve working with imperfect, noisy and sparse data. Alternatively, we could also consider the \textit{structural identifiability}, which considers the case where we have highly-idealised, noise-free observations~\cite{Chis2011}.  Structural identifiability for ODE models can be assessed using algebraic methods that are available in several software packages, such as GenSSI~\cite{GENSSI2011}, DAISY~\cite{Bellu2007} and STRIKE-GOLDD~\cite{Villaverde2016}.   
Assessing practical identifiability is a stronger condition than assessing structural identifiability because structurally identifiable models can turn out to be practically non-identifiable when working with finite, noisy data.  Previous work has established that likelihood-based methods can perform well regardless of whether a model is structurally or practically non-identifiable~\cite{Frohlich2014,Simpson2024}.

We have worked with a likelihood-based framework, which we chose for the sake of algorithmic simplicity and computational efficiency.  In particular, working with profile likelihood is often faster than sampling based methods, such as MCMC~\cite{simpson2020practical}, and this is particularly relevant for poorly identified problems. All problems we have considered here involve parameters that are well-identified by the data. {\blue We note that if the variance in the noise was larger than assumed here, or if the data were sparser, this could result in parameters being poorly identified. We anticipate that, where identifiability issues exist, they would affect both the basic and the structured methods similarly. It is possible that the structured approach might speed up the computations needed to diagnose the identifiability issue. However, we leave a detailed exploration of this for future work.}    One way of doing this would be to seek an appropriate re-parameterisation of the log-likelihood function, which we leave for future consideration. 

Our structured inference approach could, in principle, be used in a Bayesian setting by sampling from the distribution of outer parameters and, for each sample point, optimising the inner parameters using a likelihood or approximate likelihood function (see also \cite{raimundez2023posterior}). Implementing this is beyond the scope of this paper but would be an interesting aim for future work.

\subsection*{Acknowledgements} 
MJS is supported by the Australian Research Council (DP230100025).
MJP acknowledges travel support from the Australian Research Council (DP200100177). The authors thank the organisers of the 3rd New Zealand Workshop on Uncertainty Quantification and Inverse Problems, held at the University of Canterbury in 2023. The authors are grateful to five anonymous reviewers for comments on previous versions of this manuscript.

%\printbibliography
%\bibliography{references}

\end{document}

%\linenumbers

\maketitle

%\tableofcontents 
%\clearpage

\subsection*{Sensitivity analysis}

To investigate the robustness of our results around the relative performance of the basic and structured methods, we repeated the analysis in the main article with random variation in the true parameter values. Instead of generating $M=500$ datasets using fixed parameter values, we independently randomised the values of the target model parameters for each dataset. For each of the $M$ realisations, each parameter value was drawn from an independent normal distribution with the same mean as in the primary analysis (see Table \ref{tab:params}) and coefficient of variation $0.1$.

To reproduce a situation where the true parameter values are unknown, we initialised the optimisation algorithm for finding the MLE at the same fixed parameter values each time. Because this could potentially result in initial conditions that were relatively distant from the true parameter values, there was a greater risk of the optimisation algorithm to converging to a local minimum or failing to converge altogether. To mitigate this, we used the {\em GlobalSearch} function in Matlab R2022b. This algorithm tests multiple start points, generated using a scatter-search mechanism, to sample multiple basins of attraction for {\em fmincon}.

The results of the sensitivity analysis are shown in Table \ref{tab:sensitivity}. The relative errors in the MLE were slightly larger than in the primary analysis and the 95\% CI coverage rates were lower. This is possibly caused by the optimisation algorithms converging to a local minimum in some of the $M=500$ realisations. However, the accuracy and coverage properties of the structured method were comparable to the basic method for all three models, and the structured method was still substantially more efficient than the basic method. The main difference from the primary analysis was that the structured method was less efficient than the basic method for calculation of the MLE in the predator-prey, although it was still more efficient overall (for calculating MLE and parameter profiles) for all three models.

\begin{table}
\begin{tabular}{ll}
\hline
Parameter & Value \\
\hline
{\em Predator--prey model} & \\
Prey intrinsic growth rate & $r=0.5$ \\
Prey carrying capacity & $K=5000$ \\
Predation rate & $a=0.75$ \\
Saturation constant & $b=1000$ \\
Predator death rate & $\mu=0.5$ \\
Observation probability & $p_\mathrm{obs}=0.1$ \\
Initial prey population size & $R(0)=500$ \\
Initial predator population size & $P(0)=500$ \\
\hline
{\em SEIR model} & \\
Basic reproduction number  & $R_0=1.3$ \\
Latent period & $1/\gamma=2$ days \\
Infectious period & $1/\mu=4$ days \\
Immune period & $1/w=300$ days \\
Proportion of infections that are observed & $p_\mathrm{obs}=0.1$ \\
Mean time from onset of infectiousness to notification & $1/\alpha=3$ days \\
Negative binomial dispersion parameter & $k=30$ \\
Number of seed infections & $E_0=100$ \\
Population size & $N=5\times 10^6$ \\
\hline
{\em Advection--diffusion model} & \\
Diffusion coefficient & $D=1$ \\
Advection velocity & $v=0.5$ \\
Retardation factor & $R=2$ \\
Solute boundary concentration & $U_0=100$ \\
Observation time & $t_\mathrm{obs}=100$ \\
Standard deviation of observation noise & $\sigma=3$ \\
\hline
\end{tabular}
\caption{Model parameter values}
\label{tab:params}
\end{table}

\begin{table}
\centering
\small
\begin{tabular}{lllll} 
\hline  
& & \multicolumn{2}{l}{\bf Relative error (\%)} & \\ 
 & & Basic & Structured & \\ 
Predator-prey && 1.0 [0.5, 2.1] & 0.9 [0.5, 2.0] & \\ 
SEIR && 2.8 [1.3, 5.5] & 3.1 [1.4, 6.0] & \\ 
Adv. diff. && 10.7 [6.5, 17.7] & 10.7 [6.5, 17.7] & \\ 
\hline  
& & \multicolumn{2}{l}{\bf 95\% CI coverage} & \\ 
  && Basic & Structured \\ 
Predator-prey & $r$ & 81.4\% & 83.0\%  \\ 
& $a$ & 67.0\% & 68.2\%  \\ 
& $\mu$ & 67.2\% & 68.4\%  \\ 
& $p_{obs}$ & 82.4\% & - \\ 
SEIR & $R_0$ & 84.4\% & 81.6\%  \\ 
& $w$ & 84.2\% & 80.2\%  \\ 
& $p_{obs}$ & 84.0\% & - \\ 
& $k$ & 85.4\% & 82.0\%  \\ 
Adv. diff. & $D$ & 93.2\% & 93.2\%  \\ 
& $v$ & 92.2\% & 92.2\%  \\ 
& $R$ & 93.8\% & - \\ 
& $\sigma$ & 90.4\% & 90.4\%  \\ 
\hline  
&& \multicolumn{3}{l}{\bf Function calls (MLE) } \\ 
  && Basic & Structured & Improvement (\%) \\ 
Predator-prey && 4069 [3325, 5741] & 5291 [4210, 6804]  & -23.7 [-77.1, 14.9] \\ 
SEIR && 12550 [8648, 18116] & 7080 [5312, 9319]  & 42.6 [10.1, 63.2] \\ 
Adv. diff. && 11952 [8101, 18050] & 7445 [6356, 8816]  & 38.6 [7.0, 57.4] \\ 
\hline  
&& \multicolumn{3}{l}{\bf Function calls (profiles) } \\ 
  && Basic & Structured & Improvement (\%) \\ 
Predator-prey && 17665 [16469, 19199] & 9959 [8761, 12730]  & 42.8 [33.8, 48.0] \\ 
SEIR && 13323 [12407, 14597] & 7661 [7310, 8107]  & 43.1 [38.8, 46.6] \\ 
Adv. diff. && 9915 [9433, 10391] & 4628 [4193, 5007]  & 53.6 [50.2, 56.9] \\ 
\hline  
&& \multicolumn{3}{l}{\bf Function calls (total) } \\ 
  && Basic & Structured & Improvement (\%) \\ 
Predator-prey && 22330 [20391, 24860] & 16266 [13769, 19196]  & 28.1 [18.1, 37.9] \\ 
SEIR && 26527 [21852, 32200] & 14926 [13264, 17354]  & 42.2 [29.0, 53.9] \\ 
Adv. diff. && 21692 [18131, 28665] & 12132 [10893, 13492]  & 45.2 [32.2, 56.2] \\ 
\hline  
\end{tabular} 

\caption{Results from a sensitivity analysis with random variation in the true values of the target model parameters. Relative error in the maximum likelihood estimate, coverage of the 95\% CI (i.e. proportion of datasets for which the 95\% CI contained true parameter value) for each profiled parameter, and the number of calls to the forward model solver (number to calculate the MLE, number to calculate the likelihood profiles, and total number) under the basic and structured inference methods for each of the three model case studies. Note coverage statistics for the inner parameters ($p_\mathrm{obs}$ and $R$) are not applicable for the structured method, as the inner parameter is calculated as a function of the outer parameters. The `Improvement' column shows the reduction in the number of function calls under the structured method relative to the basic method. Results show the median and interquartile range across $M=500$ independently generated datasets for each model. }
\label{tab:sensitivity}
\end{table}

%\printbibliography